\documentclass{scrartcl}
\pdfoutput=1

\PassOptionsToPackage{
    numbers, 
    sort&compress, 
    super
}{natbib}
\usepackage[preprint]{modified_neurips_2023}

\usepackage[font=small,labelfont={bf}]{caption}
\usepackage{subcaption}
\usepackage{fonttable}
\usepackage[utf8]{inputenc} 
\usepackage[OT1]{fontenc}    
\usepackage[modules=all]{chemmacros}
\usepackage{upgreek}
\usepackage{xcolor}
\definecolor{TUMblue}{HTML}{005293}

\usepackage{url}            
\usepackage{booktabs}       
\usepackage{amsfonts}       
\usepackage{pifont} 
\newcommand{\cmark}{\ding{51}}%
\newcommand{\xmark}{\ding{55}}%
\usepackage{multirow}
\usepackage{nicefrac}       
\usepackage{microtype}      
\usepackage[symbol]{footmisc}
\usepackage{tablefootnote}
\usepackage{xcolor}         
\usepackage{adjustbox}
\usepackage{acro}
\acsetup{list/template=description}
\DeclareAcronym{qHTS}{
	short = qHTS,
	long = quantitative high-throughput screening,
}

\DeclareAcronym{LLM}{
	short = LLM,
	long = large language model,
}

\DeclareAcronym{MT-DNN}{
	short = MT-DNN,
	long = multi-task deep neural network,
}

\DeclareAcronym{jpg}{
	short = JPEG ,
	sort = jpeg ,
	alt = JPG ,
	long = Joint Photographic Experts Group
}

\DeclareAcronym{ML}{
	short = ML,
	long = machine learning
}

\DeclareAcronym{DL}{
	short = DL,
	long = deep learning
}

\DeclareAcronym{MCC}{
	short = MCC,
	long = Matthews correlation coefficient
}

\DeclareAcronym{Sp}{
	short = Sp,
	long = specificity
}

\DeclareAcronym{Sn}{
	short = Sn,
	long = sensitivity
}

\DeclareAcronym{BA}{
	short = BA,
	long = balanced accuracy
}

\DeclareAcronym{AP}{
	short = AP,
	long = average precision
}

\DeclareAcronym{ROC_AUC}{
	short = ROC AUC,
	long = receiver operating characteristic area under curve
}

\DeclareAcronym{PR_AUC}{
	short = PR AUC,
	long = precision recall area under curve
}

\DeclareAcronym{DPR_AUC}{
	short = $\Delta$PR AUC,
	long = $\Delta$ in precision recall area under curve
}

\DeclareAcronym{TPR}{
	short = TPR,
	long = true positive rate
}

\DeclareAcronym{TNR}{
	short = TNR,
	long = true negative rate
}

\DeclareAcronym{FPR}{
	short = FPR,
	long = false positive rate
}

\DeclareAcronym{FNR}{
	short = FNR,
	long = false negative rate
}

\DeclareAcronym{TP}{
  short = TP,
  long  = true positive
}

\DeclareAcronym{FN}{
  short = FN,
  long  = false negative
}

\DeclareAcronym{FP}{
  short = FP,
  long  = false positive
}

\DeclareAcronym{TN}{
  short = TN,
  long  = true negative
}

\DeclareAcronym{RF}{
	short = RF,
	long = random forest
}

\DeclareAcronym{AID}{
	short = AID,
	long = bioassay identifier
}

\DeclareAcronym{HTS}{
	short = HTS,
	long = high-throughput screening
}

\DeclareAcronym{MMP}{
	short = MMP,
	alt = $\Delta\Psi_\text{m}$,
	long = mitochondrial membrane potential 
}

\DeclareAcronym{m-MPI}{
	short = m-MPI,
	long = mitochondrial membrane potential indicator
}

\DeclareAcronym{ECACC}{
	short = ECACC,
	long = European Collection of Authenticated Cell Cultures 
}

\DeclareAcronym{DMEM}{
	short = DMEM,
	long = Dulbecco's modified eagle medium 
}

\DeclareAcronym{FCS}{
	short = FCS,
	long = fetal calf serum
}

\DeclareAcronym{RT}{
	short = RT,
	long = room temperature
}

\DeclareAcronym{FCCP}{
	short = FCCP,
	long = carbonylcyanid-4-(trifluormethoxy)phenylhydrazon,
}

\DeclareAcronym{DMSO}{
	short = DMSO,
	long = dimethyl sulfoxide,
}

\DeclareAcronym{ddH2O}{
	short = \ch{ddH2O},
	long = double destilled water,
	sort={ddH2O},
}

\DeclareAcronym{PBS}{
	short = PBS,
	long = phosphate-buffered saline,
}

\DeclareAcronym{EC50}{
	short = EC\textsubscript{50},
	long = half maximal effective concentration,
}

\DeclareAcronym{AI}{
	short = AI,
	long = artificial intelligence,
}

\DeclareAcronym{DTI}{
	short = DTI,
	long = drug--target interaction,
}

\DeclareAcronym{DDI}{
	short = DDI,
	long = drug--drug interaction,
}

\DeclareAcronym{DNA}{
	short = DNA,
	long = deoxyribonucleic acid,
}

\DeclareAcronym{ECFP}{
	short = ECFP,
	long = extended-connectivity fingerprint,
}

\DeclareAcronym{FCFP}{
	short = FCFP,
	long = functional-class fingerprint,
}

\DeclareAcronym{MAP4}{
	short = MAP4,
	long = MinHashed atom-pair fingerprint,
}

\DeclareAcronym{SVM}{
	short = SVM,
	long = support vector machine,
}

\DeclareAcronym{DNN}{
	short = DNN,
	long = deep neural network,
}

\DeclareAcronym{GCNN}{
	short = GCNN,
	long = graph convolutional neural networks,
}

\DeclareAcronym{GHS}{
	short = GHS,
	long = globally harmonized system of classification and labelling of chemicals,
}

\DeclareAcronym{SMILES}{
	short = SMILES,
	long = simplified molecular-input line-entry system,
}

\DeclareAcronym{CLI}{
	short = CLI,
	long = command-line interpreter,
}

\DeclareAcronym{GUI}{
	short = GUI,
	long = graphical user interface,
}

\DeclareAcronym{ATP}{
	short = ATP,
	long = adenosine triphosphate,
}

\DeclareAcronym{AMP}{
	short = AMP,
	long = adenosine monophosphate,
}

\DeclareAcronym{PPi}{
	short = PP\textsubscript{i},
	long = pyrophosphate,
	sort={PPi}
}

\DeclareAcronym{Lu}{
	short = \ch{LH2},
	long = luciferin,
	sort={LH2}
}

\DeclareAcronym{oLu}{
	short = \ch{oxy-L},
	long = oxy-luciferin,
	sort={oxylu}
}

\DeclareAcronym{SMOTE}{
	short = SMOTE,
	long = synthetic minority over-sampling technique,
}

\DeclareAcronym{SHAP}{
	short = SHAP,
	long = Shapley additive explanation,
}

\DeclareAcronym{CPU}{
	short = CPU,
	long = central processing unit,
}

\DeclareAcronym{RAM}{
	short = RAM,
	long = random-access memory,
}

\DeclareAcronym{GPU}{
	short = GPU,
	long = graphics processing unit,
}

\DeclareAcronym{CAS}{
	short = CAS,
	long = Chemical Abstracts Service,
}

\DeclareAcronym{UMAP}{
	short = UMAP,
	long = uniform manifold approximation and projection,
}

\DeclareAcronym{Tox21}{
	short = Tox21,
	long = Toxicology in the 21\textsuperscript{st} Century,
}

\DeclareAcronym{GOSS}{
	short = GOSS,
	long = gradient-based one-side sampling,
}

\DeclareAcronym{QSAR}{
	short = QSAR,
	long = quantitative structure--activity relationship,
}

\DeclareAcronym{EFB}{
	short = EFB,
	long = exclusive feature bundling,
}

\DeclareAcronym{MTT}{
	short = MTT,
	long = \iupac{3-(4, 5-dimethylthiazolyl-2)-2,5-diphenyltetrazolium bromide},
}

\DeclareAcronym{SSL}{
	short = SSL,
	long = self-supervised learning,
}

\DeclareAcronym{GBM}{
	short = GBM,
	long = gradient boosting machine,
}

\DeclareAcronym{MLP}{
	short = MLP,
	long = multilayer perceptron,
}

\DeclareAcronym{API}{
	short = API,
	long = application programming interface,
}

\DeclareAcronym{CHOP}{
  short = CHOP,
  long  = C/EBP homologous protein
}
\DeclareAcronym{UPR}{
  short = UPR,
  long  = unfolded protein response
}
\DeclareAcronym{ER}{
  short = ER,
  long  = endoplasmic reticulum
}

\DeclareAcronym{PN}{
  short = PN,
  long  = prototypical network
}

\DeclareAcronym{FH}{
  short = FH,
  long  = frequent hitters
}

\DeclareAcronym{LSA}{
  short = LSA,
  long  = latent semantic analysis
}

\usepackage{siunitx}
\usepackage{physics}
\usepackage{lipsum}
\usepackage{csquotes}
\makeatletter
\renewcommand{\mkbegdispquote}[2]{%
    \leavevmode\llap{\openautoquote\csq@eqgroup}%
    \csq@bqgroup
    \advance\csq@qlevel\@ne
    \csq@resetstyle
    \csq@init}
    
\renewcommand\mkblockquote[4]{%
    \leavevmode\llap{\openautoquote\csq@eqgroup}%
    \csq@bqgroup
    \advance\csq@qlevel\@ne
    \csq@resetstyle
    \csq@init
    #1#2#3\closeautoquote#4}
\makeatother

\usepackage{orcidlink} 
\hypersetup{
    colorlinks=true,
    allcolors=TUMblue
}
\usepackage{cleveref}

\AtBeginDocument{\RenewCommandCopy\qty\SI}

\title{\textsc{TwinBooster:} Synergising Large Language Models with Barlow Twins and Gradient Boosting for Enhanced Molecular Property Prediction}

\author{%
  Maximilian G. Schuh\,\orcidlink{0009-0008-2415-8810}\\
    \texttt{\href{mailto:m.schuh@tum.de}{m.schuh@tum.de}}\\
  \And
  Davide Boldini\,\orcidlink{0000-0002-4109-3528}\\
    \texttt{\href{mailto:davide.boldini@tum.de}{davide.boldini@tum.de}} \\
  \AND
  Stephan A. Sieber\,\orcidlink{0000-0002-9400-906X}\\
  Chair of Organic Chemistry II\\
  TUM School of Natural Sciences\\
  Technical University of Munich\\
    \texttt{\href{mailto:stephan.sieber@tum.de}{stephan.sieber@tum.de}} \\
}

\begin{document}

\maketitle

\begin{abstract}
The success of drug discovery and development relies on the precise prediction of molecular activities and properties. 
While \textit{in silico} molecular property prediction has shown remarkable potential, its use has been limited so far to assays for which large amounts of data are available.
In this study, we use a fine-tuned \acl{LLM} to integrate biological assays based on their textual information, coupled with Barlow Twins, a Siamese neural network using a novel \acl{SSL} approach. 
This architecture uses both assay information and molecular fingerprints to extract the true molecular information.
\textsc{TwinBooster} enables the prediction of properties of unseen bioassays and molecules by providing state-of-the-art zero-shot learning tasks.
Remarkably, our \acl{AI} pipeline shows excellent performance on the FS-Mol benchmark.
This breakthrough demonstrates the application of \acl{DL} to critical property prediction tasks where data is typically scarce. 
By accelerating the early identification of active molecules in drug discovery and development, this method has the potential to help streamline the identification of novel therapeutics.
\end{abstract}

\acresetall

\section{Introduction}
Accurate prediction of biomolecular properties, such as toxicity,\cite{zhang2019lightgbm} is a critical factor in accelerating the drug discovery and development process.\cite{deng2023systematica,chithrananda2020chemberta,jain2021largescale,walter2022analysisa}
However, the reliance on traditional laboratory experiments presents significant challenges. 
These methods are not only time-consuming and expensive, but resource constraints make them impractical when scaled up to large numbers of molecules.\cite{yang2019analyzing,shen2019molecular}

To bridge this gap and improve predictive accuracy, it is essential to collect a significant amount of data. 
In biomolecular research, the quantity and quality of data points are critical to the development of robust predictive models. 
Without a large dataset, models may lack the precision and reliability needed to identify potential drug candidates and assess their safety profiles.\cite{shen2019molecular,gaulton2012chembl}

To address these challenges, the \textit{in silico} analysis of chemical structures and bioassays is emerging as a promising solution.\cite{jain2021largescale}
This computational approach uses large data sets to train more effective predictive models.\cite{seidl2023enhancing}
By virtual modelling experiments, it bypasses the limitations of traditional lab-based methods and offers a faster, more cost-effective and scalable alternative for studying a wide range of biomolecules. 
This innovative method not only streamlines the drug development process, but also enhances the predictive capabilities critical to identifying viable drug candidates.\cite{merkwirth2005automatic}

Advances in \ac{LLM} technology are opening up new ways of reinterpreting large datasets, particularly in the field of bioassays.\cite{young2018recent}
Our research exploits this potential by fine-tuning an \ac{LLM} specifically for the task of integrating and understanding textual information from assay titles, descriptions and protocols to predict molecular properties.\cite{young2018recent}
This approach, which is unique in its application, leverages PubChem's comprehensive data repository of over \num{1500000} bioassays.\cite{kim2023pubchem}

Our method applies a fine-tuned \ac{LLM} to accurately capture and interpret the semantic nuances of bioassay text.\cite{he2021deberta,he2023debertav3} 
The \ac{LLM} extracts and integrates complex information to generate meaningful semantic embeddings. 
This advanced capability enhances the depth and quality of our molecular property predictions, providing a novel and effective way to analyse bioassay data.
    
\begin{figure}[!htb]
    \centering
    \includegraphics[width=.9\textwidth]{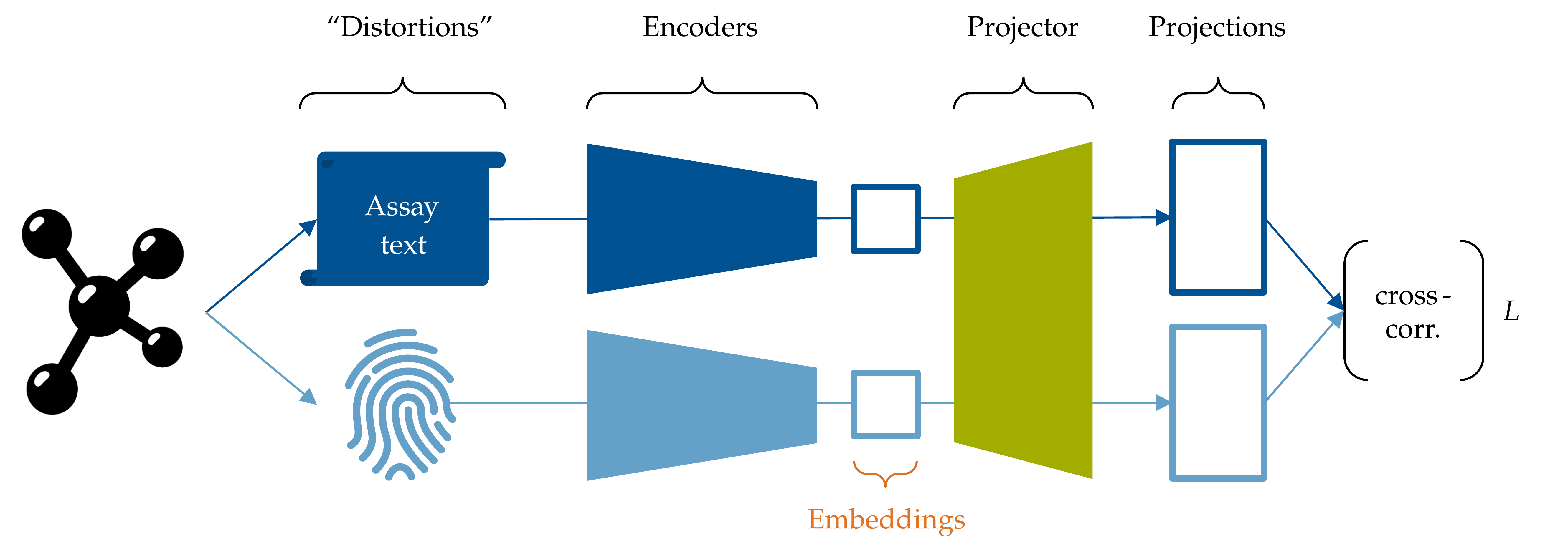}
    \caption{The \textsc{TwinBooster} architecture. This Siamese neural network provides an information-rich and bias-free representation of molecules in the context of bioassays.\cite{zbontar2021barlow}}
    \label{fig:bt_arch}
\end{figure}

In our pursuit of enhancing \ac{QSAR}, we introduce \textsc{TwinBooster}, a classification architecture inspired by Barlow Twins.\cite{zbontar2021barlow} 
The primary advantage of Barlow Twins over other \ac{SSL} techniques lies in its novel objective function, which measures the cross-correlation matrix between the outputs of two identical networks processing different representations of a molecule. 
We are using \acp{ECFP} and the corresponding bioassay text embedded by the fine-tuned \ac{LLM}.\cite{landrum2020rdkit,he2023debertav3,he2021deberta}
The aim is to make this matrix as close as possible to the identity matrix. 
This approach not only ensures the similarity of the embedding vectors for distorted versions of a molecule, but also minimises the redundancy between the components of these vectors, thereby revealing a representation that is rich in information and free of bias (shown in \cref{fig:bt_arch}).\cite{hadsell2006dimensionality}
Notably, Barlow Twins does not require a large number of negative samples, allowing it to work effectively on smaller batches. 
It also performs better on very high-dimensional embeddings compared to current methods.\cite{zbontar2021barlow}

\Ac{QSAR} modelling is essential in cheminformatics research, enabling \textit{in silico} predictions of molecular properties.
With information-rich representations generated by the Barlow Twins architecture decision tree ensembles such as \acp{GBM} are used in this study due to their remarkable performance, ability to rank features, and scalability.\cite{biau2016random,jiang2021could,boldini2023practical,stanley2021fsmol}
In recent years, \acp{GBM} have become increasingly popular in cheminformatics for a range of tasks, such as predicting toxicity, analysing drug sensitivity, modelling anti-cancer activity, and identifying drug-target interactions.\cite{zhang2019lightgbm,breiman2017classification}
\acp{GBM} are able to tackle broad ranges of dataset sizes and class-imbalance ratios, ideal for scenarios in drug discovery and development applications.\cite{zheng2021drugcomb,winter2019learning}
Combining \acp{GBM} with the information-rich representation provided by the Barlow Twins architecture results in state-of-the-art performance in the zero-shot classification task.
To enhance the robustness and predictive power of our model, we move from a conventional zero-shot framework to a novel pseudo-proteochemometric approach. 
Here, a \ac{GBM} is trained on the information bottleneck embeddings\cite{tishby2015deep} derived from the Barlow Twins architecture.\cite{zbontar2021barlow,xian2020zeroshot} 
This strategic shift enables the \ac{GBM} to operate effectively in zero-shot tasks, where its predictive capabilities are tested on bioassays beyond the scope of its training dataset. 
These assays, which are new to the model, encompass previously unseen biological targets and assay types, providing a rigorous test of the model's ability to capture and analyse previously unseen data.

In conclusion, \textsc{TwinBooster} and this study contribute significantly to drug discovery and development through:
\begin{enumerate}
    \item Achieve state-of-the-art performance in zero-shot classification tasks, critical for drug discovery pre-screening.

    \item Provide an intuitive user experience for experimentalists in molecular property prediction using \ac{ML}, \ac{DL} and \ac{LLM} technology, enabling faster and more cost-effective drug discovery.

    \item Present a conformal prediction implementation that assesses the confidence of molecular property predictions.

    \item Present a case study of an \textit{in silico} pre-screening experiment, emphasising the design of experiments for increased efficiency and higher chances of discovering desired hits.
\end{enumerate}

\section{Materials and Methods}

\subsection{Dataset}

\paragraph{FS-Mol} The FS-Mol dataset\cite{stanley2021fsmol} proposes a new approach to drug discovery using few-shot learning, to analyse small datasets, which are common in drug discovery due to high data generation costs and ethical considerations. 
The classification dataset and benchmarking procedure are designed to simulate the challenges of machine learning in drug discovery, where typically only a few hundred compounds can be tested. 
FS-Mol evaluates single-task, multi-task and meta-learning approaches and contains \ac{ML} baselines.
It provides training, validation as well as testing data, which are sourced from ChEMBL.\cite{mendez2019chembl}
In the context of few-shot learning a set from 16 up to 256 support molecules, alongside binary activity labels are provided.\cite{stanley2021fsmol}

\subsubsection{Molecular representation}
\paragraph{Bioassay-based \ac{LLM} text embeddings}
The pipeline in this study requires titles, descriptions and protocols as additional representation for each molecule and assay. 
Therefore, this text information is extracted from PubChem.\cite{kim2023pubchem}
Using both \Acp{API} from PubChem and ChEMBL a mapping of \ac{AID} to ChEMBL IDs is performed.\cite{kim2023pubchem,mendez2019chembl}
This is done to retrieve the information rich text information of PubChem in combination with the ChEMBL-based FS-Mol benchmark.

Finally, the text is converted into a vector (of shape 768) using our fine-tuned \ac{LLM} PubChemDeBERTa.

\paragraph{\Aclp{ECFP}}
All molecules are handled in \ac{SMILES} strings then converted to \acp{ECFP} 1024 bits and a radius of 2, using the Python\cite{vanrossum1995python} Rdkit\cite{landrum2020rdkit} implementation.

\subsection{Models}

\subsubsection{\Acl{LLM}}

\paragraph{Fine-tuning}
The DeBERTa V3 base model\cite{he2023debertav3} is fine-tuned on the PubChem corpus using \SI{\sim14}{\giga\byte} video \ac{RAM} for \SI{\sim15}{\hour}.
In the augmented version, the description is shuffled (\enquote{.} as delimiter) and 5 augmentations are used as the training corpus.
Therefore, the Python\cite{vanrossum1995python} Transformers\cite{wolf2020transformers} library is used.
The Optuna\cite{akiba2019optuna} hyperparameter optimisation library is used to find the best combination of hyperparameters for the \ac{LLM} (ref. \cref{tab:llm_hyperparameter_space}).
After 20 optimisation procedure trials the best hyperparameters shown in \cref{tab:llm_hyperparameters} were found.

\begin{table}[!htb]
\centering
\caption{\ac{LLM} hyperparameter optimisation space.}
\label{tab:llm_hyperparameter_space}
\begin{tabular}{ll}
\toprule
\textbf{Hyperparameter} & \textbf{Range} \\
\midrule
learning\_rate & \{\num{1.5e-5}, \num{2e-5}, \num{2.5e-5}, \num{3e-5}\} \\
batch\_size & \{16, 32\} \\
max\_length & \{64, 128\} \\
num\_train\_epochs & 1.0 \\
\bottomrule
\end{tabular}
\end{table}

\begin{table}[!htb]
\centering
\caption{Best \ac{LLM} fine-tuning hyperparameters.}
\label{tab:llm_hyperparameters}
\begin{tabular}{lS}
\toprule
\textbf{Hyperparameter} & \textbf{Value} \\
\midrule
ampere & \text{True} \\
num\_train\_epochs & 3.0 \\
learning\_rate & 3e-5 \\
weight\_decay & 0.01 \\
batch\_size & 32 \\
max\_length & 128 \\
adam\_beta1 & 0.9 \\
adam\_beta2 & 0.999 \\
adam\_epsilon & 1e-6 \\
warmup\_steps & 500 \\
\bottomrule
\end{tabular}
\end{table}

\paragraph{Performance evaluation} 
For the evaluation, the perplexity is used as an evaluation metric for fine-tuning the \ac{LLM} ($\mathrm{perplexity} \in [0, \infty)$, lower values indicate better performance).\cite{jelinek1977perplexity}
The \ac{LLM} evaluation was performed on the complete training corpus with a token masking rate of \SI{15}{\percent}.
Since our investigation focuses exclusively on the \ac{LLM} behaviour in-distribution and not out-of-distribution, other performance metrics are not considered or evaluated.\cite{meister2021language}

The training and evaluation procedures are conducted using the PubChem corpus.\cite{kim2023pubchem} 
This corpus is selected to generate optimal embeddings that represent the bioassays for our model.

\subsubsection{\textsc{TwinBooster}}

\paragraph{\Acl{MLP}} Barlow Twins use \acfp{MLP} for both the encoders and the projector design. 
The network architecture is altered from the original by having two encoders a molecule and a text encoder.
Finally, the projector is shared for both representations.

Both encoders as well as the projector have the following structure
\begin{equation*}
    \boldsymbol{l_{i+1}} = \mathrm{Linear} \left( {\phi \left( \mathrm{BatchNorm} \left( \mathrm{Linear} \left(\boldsymbol{W l_i + b} \right) \right) \right)}^{n} \right),
\end{equation*} 

where $\boldsymbol{l_{i}}$ is the input layer and $\boldsymbol{l_{i+1}}$ is its output, with a flexible number of layers $n$ and adjustable dimensionality of input and output. 
Furthermore, variables $\boldsymbol{W}$, $\boldsymbol{b}$ represent learnable weights and biases. 
A linear layer is followed by batch normalisation,\cite{ioffe2015batch} an activation function $\phi$,\cite{agarap2019deep,ramachandran2017searching} and the last linear layer.
The network is constructed using PyTorch.\cite{paszke2019pytorch}

For training the network is using Barlow Twins loss\cite{zbontar2021barlow} and the AdamW optimiser.\cite{loshchilov2019decoupled} 
Manual hyperparameter tuning was performed on a range and set of parameters listed in \cref{tab:bt_hyperparameter_space}.
The model is trained for 25 epochs or until early stopping was engaged if a validation set is provided.

Furthermore, the model is trained using \acp{ECFP} and \ac{LLM} embeddings. 
For inactive molecules, the embeddings are sign changed.

\begin{table}[!htb]
\centering
\caption{Barlow Twins Hyperparameters. The range of parameters is listed and the best are highlighted in bold.}
\begin{tabular}{ll}
\toprule
\textbf{Hyperparameter} & \textbf{Range} \\
\midrule
enc\_n\_neurons & \{512, 1024, 2048, \textbf{4096}, 8192\} \\
enc\_n\_layers & \{2, 3, \textbf{4}\} \\
proj\_n\_neurons & \{512, 1024, \textbf{2048}, 4096, 8192\} \\
proj\_n\_layers & \{\textbf{2}, 3, 4\} \\
embedding\_dim & \{512, \textbf{1024}, 2048, 4096\} \\
act\_function & \{ReLU,\cite{agarap2019deep} \textbf{Swish}\cite{ramachandran2017searching}\} \\
batch\_size & \{\textbf{1024}, 2048\} \\
learning\_rate & $\{\num{5e-3}, \textbf{\num{1e-4}}\}$ \\
weight\_decay & $\{\num{1e-3}, \textbf{\num{5e-3}}\}$ \\
\bottomrule
\end{tabular}
\label{tab:bt_hyperparameter_space}
\end{table}

\paragraph{\Acl{GBM}} \label{sec:gbm_hp}
The \ac{GBM} package LightGBM is used for for training based on the informational bottleneck embeddings provided by the Barlow Twins model.\cite{tishby2015deep}
Performing zero-shot predictions is done by feeding the Barlow Twins model with \acp{ECFP} and text information of the desired molecules.\cite{ke2017lightgbm}
To achieve optimal performance, the SMAC3\cite{lindauer2022smac3} hyperparameter optimisation library is applied to find the optimal combination of hyperparameters (ref. \cref{tab:gbm_hyperparameter_space}), using \SI{80}{\percent} of the \enquote{train} data of the FS-Mol dataset for training.
Optimisation is set to 200 trials.
The evaluation is performed by assessing the \ac{PR_AUC} and \ac{ROC_AUC} on the \enquote{valid} and the remaining \SI{20}{\percent} of the \enquote{train} data of the FS-Mol benchmark. 
SMAC3's multi-fidelity implementation is used with the budget parameter being represented by the n\_estimators parameter of LightGBM.\cite{lindauer2022smac3,ke2017lightgbm}

\begin{table}[!htb]
\centering
\caption{\Ac{GBM} SMAC3 hyperparameter optimisation space.}
\label{tab:gbm_hyperparameter_space}
\begin{tabular}{ll}
\toprule
\textbf{Hyperparameter} & \textbf{Range} \\
\midrule
budget (n\_estimators) & $[200, 2000]$ \\
\midrule
num\_leaves & $[62,256]$ (step size 64) \\
learning\_rate & $[\num{1e-8},1.0]$ (log scale) \\
min\_child\_samples & $[5,100]$ \\
subsample & $[0.4,1.0]$ \\
subsample\_freq & $[0,7]$ \\
reg\_lambda & $[\num{1e-8},10.0]$ \\
\bottomrule
\end{tabular}
\end{table}

Finally, the LightGBM is trained using the full \enquote{train} data of the FS-Mol dataset and the best hyperparameters listed in \cref{tab:gbm_hyperparameter}.

\begin{table}[!htb]
\centering
\caption{Best \ac{GBM} hyperparameters after optimisation.}
\label{tab:gbm_hyperparameter}
\begin{tabular}{lS}
\toprule
\textbf{Hyperparameter} & \textbf{Value} \\
\midrule
budget (n\_estimators) & 2000 \\
\midrule
num\_leaves & 256 \\
learning\_rate & 0.0711 \\
min\_child\_samples & 60 \\
subsample & 0.941 \\
subsample\_freq & 1 \\
reg\_lambda & 3.78 \\
\bottomrule
\end{tabular}
\end{table}

\paragraph{Performance evaluation} When comparing models, we are using intersecting tasks of the \enquote{test} data, to ensure a scientific comparison. 
Metric selection is based on the FS-Mol benchmark.\cite{stanley2021fsmol} 
\Acp{ROC_AUC} are commonly used for classifier evaluation in the presence of class imbalance, but they can be less reliable for rare classes due to small sample sizes.\cite{fawcett2006introduction,tharwat2020classification}

\begin{equation*}
	\label{eq:ROCAUC}
	\begin{aligned}
		\text{\acs{ROC_AUC}} &= \int_{0}^1 f_\text{\acs{TPR}}\left(f_\text{\acs{FPR}}\right) \dd f_\text{\acs{FPR}} \\
		\mathrm{TPR} &= \dfrac{\mathrm{TP}}{\mathrm{TP}+\mathrm{FN}} \\
        \mathrm{FPR} &= \dfrac{\mathrm{FP}}{\mathrm{FP}+\mathrm{TN}}
	\end{aligned}
\end{equation*}

\Acp{PR_AUC} are recommended for highly skewed classes, as they provide a more realistic view of classifier performance than \acp{ROC_AUC}.\cite{2018learning,branco2015survey,piryonesi2020data,tharwat2020classification}
Moreover, both metrics can be calculated based on the probability of the prediction rather than the prediction itself, where a classification threshold problem can arise.\cite{tharwat2020classification}
These metrics are also used in the FS-Mol benchmark.\cite{stanley2021fsmol}

\begin{equation*}
	\label{eq:PRAUC}
    \begin{aligned}
        \text{\acs{PR_AUC}} &= \int_0^1 f_\text{Precision}\left(f_\text{Recall}\right) \dd f_\text{Recall} \\
        \text{Precision} &= \dfrac{\mathrm{TP}}{\mathrm{TP}+\mathrm{FP}} \\
        \text{Recall} &= \dfrac{\mathrm{TP}}{\mathrm{TP}+\mathrm{FN}}
    \end{aligned}
\end{equation*}

In the context of zero and few-shot learning, a different form of \ac{PR_AUC}, known as \ac{DPR_AUC}, is used. 
Here $t_i$ denotes a particular task or bioassay within the total set of $i$ tasks. 
The expression $\sum t_i$ represents the sum of all activity endpoints for a given task, indicating the number of active molecules. 
In addition, $|t_i|$ corresponds to the size of the task. 
This metric shows sensitivity to the balance between classes, allowing for straightforward comparisons with a baseline benchmark. This is due to the performance of a random classifier reflecting the percentage of positive endpoints.\cite{stanley2021fsmol}

\begin{equation*}
    \displaystyle
    \text{\acs{DPR_AUC}}\left( t_i \right) = \text{\acs{PR_AUC}}\left( t_i \right) - \frac{\sum t_i}{|t_i|}
\end{equation*}

\paragraph{Conformal prediction} In our study, we apply the conformal prediction method using the LightGBM classifier.\cite{ke2017lightgbm,cortes-ciriano2019concepts}
This technique involves a two-step process: calibration with cross-validation on training data (5 fold), because no calibration set is provided, and prediction on test data.\cite{stone1974crossvalidatory}
Then \ac{GBM} predictions are analysed while the confidence level is set to $\epsilon = 0.80$. 
This method is valuable in providing both predictive outputs and insights into the certainty of each prediction.\cite{cortes-ciriano2019concepts}

\subsection{Case study}

To highlight the zero-shot capabilities of \textsc{TwinBooster} a case study of biological \ac{HTS} was conducted. 
Therefore, the primary screen (\ac{AID} 2732\footnote{\url{https://pubchem.ncbi.nlm.nih.gov/bioassay/2732}}) is analysed by \textsc{TwinBooster} to predict the desired properties.
Then it is analysed against the confirmatory screen (\ac{AID} 504437\footnote{\url{https://pubchem.ncbi.nlm.nih.gov/bioassay/504437}}).
The columns \texttt{PUBCHEM\_EXT\_DATASOURCE\_SMILES}, \texttt{PUBCHEM\_ACTIVITY\_OUTCOME} as well as the text information are used for the \textsc{TwinBooster} prediction pipeline.\cite{kim2023pubchem}

\paragraph{Performance evaluation} Recall measures the proportion of relevant instances that are retrieved, this refers to the active compounds in this case study.\cite{tharwat2020classification}
It can be expressed at the ratio of the found active molecules (\acp{TP}) and all active molecules (\acp{TP} and \acp{FN}).

\paragraph{Similarity estimation} The Tanimoto similarity of compounds is calculated using the corresponding Rdkit function.\cite{landrum2020rdkit}
To highlight structural similarities and differences, 50 compounds are randomly selected for visual reasons.

\section{Results and Discussion}

\paragraph{Fine-tuned \ac{LLM} on bioassay corpus}

In this study, we used Microsoft's developed DeBERTa V3\cite{he2023debertav3} as the underlying architecture for our \ac{LLM} and fine-tuned it on a comprehensive bioassay corpus obtained from PubChem.\cite{he2021deberta,he2023debertav3,kim2023pubchem}
We chose DeBERTaV3, a pre-trained \ac{LLM}, for this research due to its superior performance compared to the original DeBERTa or BERT model, respectivly. 
DeBERTa V3 uses the replaced token detection pre-training task, which is more sample-efficient than the traditional masked \ac{LLM} approach. 
This innovation enhances both training efficiency and model quality by removing the \enquote{tug-of-war} dynamics present in the vanilla embedding sharing method used in ELECTRA.\cite{he2021deberta,he2023debertav3}
The fine-tuning process aimed to enhance the model's performance in predicting biomolecular properties.

Through the fine-tuning of DeBERTa V3, a remarkable reduction in average perplexity is achieved, decreasing from \num{10.7e6} to \num{1.52} (refer to \cref{tab:llm_metrics}).\cite{he2021deberta,he2023debertav3}
In conclusion, the performance is unmatched by other \acp{LLM} like BERT or BioBERT.\cite{devlin2019bert,lee2020biobert}
This improvement indicates that the \ac{LLM} has gained a deeper understanding of the data, resulting in more accurate predictions and a better fit to our specific task, by understanding terminology like cell lines, technical equipment as well as chemicals.
This ability represents a critical foundation for further analyses.

\begin{table}[!htb]
\centering
\caption{\ac{LLM} performance evaluation. The best value is highlighted in bold.}
\label{tab:llm_metrics}
\begin{tabular}{lS[detect-weight]}
\toprule
\textbf{Model} & \textbf{Perplexity} \\
\midrule
BERT base uncased\cite{devlin2019bert} & 14.9 \\
DeBERTa base\cite{he2021deberta} & 12.5e4 \\
DeBERTa V3 base\cite{he2023debertav3} & 10.7e6 \\
BioBERT V1.2 base cased\cite{lee2020biobert} & 4.47 \\
PubChemDeBERTa & 2.32 \\
PubChemDeBERTa augmented & \bfseries 1.52 \\
\bottomrule
\end{tabular}
\end{table}

The chosen evaluation metric, perplexity, provides a reliable measure of the performance of the \ac{LLM}.
Its use is well established in the field of language modelling.\cite{jelinek1977perplexity} 
Regarding the choice of performance metrics, the primary goal of this analysis is to analyse the behaviour of the \ac{LLM} within its known data distribution. 
We are not concerned with its performance on unseen (out-of-distribution) data. 
As a result, other evaluation metrics that typically assess generalisation to out-of-distribution scenarios are not necessary for the scope of this investigation.\cite{meister2021language}

\begin{figure}[!htb]
    \centering
    \includegraphics[width=.7\textwidth]{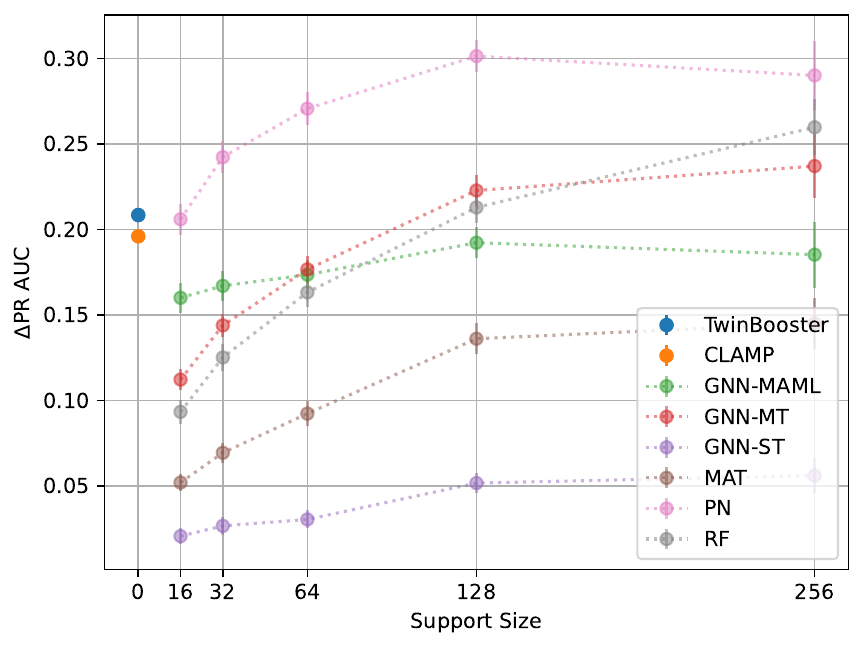}
    \caption{Zero- and few-shot FS-Mol benchmark performance of various \ac{ML}/\ac{DL} models.\cite{stanley2021fsmol,seidl2023enhancing} Standard deviations are shown between replicates.}
    \label{fig:zero_shot}
\end{figure}

\paragraph{Zero-shot benchmark}
The FS-Mol benchmark is used as the standard of measurement when assessing zero-shot capabilities, where the aim is to predict unseen tasks.
Our approach demonstrated strong performance, achieving a \ac{DPR_AUC} of \SI{20.84(24)}{\percent}, as shown in both \cref{fig:zero_shot,tab:zero_shot}. 
For a more detailed analysis, comparisons are made against two baselines: the zero-shot algorithm CLAMP,\cite{seidl2023enhancing} which reported a \ac{DPR_AUC} of \SI{19.37 \pm 0.2}{\percent}, and the few-shot learning approach \acp{PN} of FS-Mol.\cite{stanley2021fsmol}
\textsc{TwinBooster} in zero-shot learning outperforms the best few-shot baseline of the FS-Mol benchmark, \acp{PN}, at 16 support molecules with a \ac{DPR_AUC} of \SI{20.17(8)}{\percent} (Wilcoxon\cite{virtanen2020scipy} test $\alpha = 0.05$).\cite{stanley2021fsmol}

\begin{table}[!htb]
    \centering
    \caption{Comparing different zero- and few-shot model performances across different metrics on FS-Mol. In zero-shot mode no \enquote{test} molecules are provided, in the case of the few-shot performance of \acs{PN} 16 molecules of the \enquote{test} set are provided. 10 replicates each are performed. Results that are both the best and statistically significant (Wilcoxon\cite{virtanen2020scipy} test $\alpha = 0.05$) are highlighted in bold.}
    \label{tab:zero_shot}
    \sisetup{
        table-alignment-mode = none,
        table-number-alignment = center,
        table-format=.2(1),
        table-auto-round
    }
    \begin{tabular}{rSSS}
    \toprule
     & \textbf{\textsc{TwinBooster}} & \textbf{CLAMP\tablefootnote{It is not possible to make direct comparisons as only mean values and standard deviations are provided.}}\cite{seidl2023enhancing} & \textbf{\acs{PN}}\cite{stanley2021fsmol}\\
    \midrule
    Mode & \text{zero-shot} & \text{zero-shot} & \text{few-shot (16)} \\
    \midrule
    \acs{ROC_AUC} (\si{\percent}) & 71.11(29) & 69.26(20) & \text{---} \\
    \acs{PR_AUC} (\si{\percent}) & \bfseries 68.56(24) & 66.55(20) & 67.72(8)\\
    \acs{DPR_AUC} (\si{\percent}) & \bfseries 20.84(24) & 19.37(20) & 20.17(8)\\
    \bottomrule
    \end{tabular}
\end{table}

In addition, performance is assessed on confident predictions, which are evaluated using conformal prediction.
Across all metrics, performance could be significantly improved, e.g. with a relative \ac{DPR_AUC} increase of \SI{\sim10}{\percent} (shown in \cref{tab:cp_zero_shot}).
The average ratio of confident predictions across all bioassays is \SI{65}{\percent}.

\begin{table}[!htb]
    \centering
    \caption{Comparing zero-shot performances with or without conformal prediction on FS-Mol. The confidence level is set to $\epsilon = 0.80$. 10 replicates each are performed. Results that are both the best and statistically significant (Wilcoxon\cite{virtanen2020scipy} test $\alpha = 0.05$) are highlighted in bold.}
    \label{tab:cp_zero_shot}
    \sisetup{
        table-alignment-mode = none,
        table-number-alignment = center,
        table-format=.2(1),
        table-auto-round
    }
    \begin{tabular}{rSS}
    \toprule
    \multirow{2}{*}{\textbf{\textsc{TwinBooster}}} & \multicolumn{2}{c}{\textbf{Conformal Prediction}} \\
    & \xmark & \cmark \\
    \midrule
    \acs{ROC_AUC} (\si{\percent}) & 71.11(29) & \bfseries 73.76(30) \\
    \acs{PR_AUC} (\si{\percent}) & 68.56(24) & \bfseries 71.04(31) \\
    \acs{DPR_AUC} (\si{\percent}) & 20.84(24) & \bfseries 22.81(30) \\
    \bottomrule
    \end{tabular}
\end{table}

The improved performance is due to the fine-tuned \ac{LLM}, which effectively transforms bioassay text into numerical data, surpassing the capabilities of \ac{LSA}.\cite{deerwester1990indexing} 
In addition, the Barlow Twins method produces superior embeddings that capture both bioassay and molecular data.\cite{he2023debertav3,zbontar2021barlow} 
In the \ac{SSL} framework, the Barlow Twins architecture uses the information bottleneck principle\cite{tishby2015deep} to optimise representations, maximising molecular information while minimising extraneous details from \acp{ECFP} and \ac{LLM} text embeddings. 
This approach focuses on preserving important molecular details and reducing noise.\cite{zbontar2021barlow} 
Furthermore, the Barlow Twins model trains on negative as well as positive examples, which should help generalisation.
Finally, the use of a \ac{GBM} in zero-shot inference provides fast, efficient and powerful results in predictive drug discovery.\cite{boldini2023practical,ke2017lightgbm,zhang2019lightgbm}

\subparagraph{Ablation study} The ablation study is carried out to identify the differential effects on FS-Mol performance metrics attributable to each ablation.
The first step in this exploration involves the combination of \acp{ECFP} and PubChemDeBERTa bioassay embeddings for each molecule under investigation. 
The \ac{GBM} is then trained using a methodology analogous to that used in \textsc{TwinBooster}, which involves hyperparameter optimisation. 
Subsequently, the experiment is repeated with one modification: the original text embeddings are replaced by \ac{LSA} embeddings. 
These substitutions aim at evaluating the comparative effectiveness of different text embedding techniques as well as using the Barlow Twins architecture in the context of the \ac{GBM} framework.
The results of these investigations are shown in \cref{tab:ablations}.

\begin{table}[!htb]
    \centering
    \caption{Performance of different ablation experiments on FS-Mol. All results are tested pairwise using the Wilcoxon test with Bonferroni correction.\cite{virtanen2020scipy,bonferroni1936teoria} 10 replicates each are performed. Significance in Wilcoxon\cite{virtanen2020scipy} test is indicated at $\alpha = \frac{0.05}{3}$ for all results except those in italics, which are not significant. Results that are both the best and statistically significant are highlighted in bold.}
    \label{tab:ablations}
    \sisetup{
        table-alignment-mode = none,
        table-number-alignment = center,
        table-format=.2(1),
        table-auto-round,
        detect-all = true
    }
    \begin{tabular}{rSSS}
    \toprule
     & \textbf{\textsc{TwinBooster}} & \textbf{\text{ECFP + PubChemDeBERTa}} & \textbf{\text{ECFP + LSA}} \\
    \midrule
    \acs{ROC_AUC} (\si{\percent}) & \itshape 71.11 (0.29) & \itshape 70.88 (0.27) & 70.20 (0.22) \\
    \acs{PR_AUC} (\si{\percent}) & \bfseries 68.57 (0.24) & 68.13 (0.29) & 67.51 (0.21) \\
    \acs{DPR_AUC} (\si{\percent}) & \bfseries 20.84 (0.24) & 20.41 (0.29) & 19.78 (0.21) \\
    \bottomrule
    \end{tabular}
\end{table}

The ablation results show a significant improvement when using PubChemDeBERTa embeddings as opposed to \ac{LSA} embeddings. 
In addition, the use of embeddings derived from the Barlow Twins architecture as implemented in \textsc{TwinBooster} shows significant performance improvements, surpassing the combined use of \acp{ECFP} with text embeddings. 

\paragraph{Zero-shot case study} In the study conducted by \citeauthor{flaherty2014discovery}, the research team used \acp{HTS} to identify compounds that selectively activate the \ac{CHOP} pathway in the context of \ac{ER} stress.
The \ac{UPR} is a cellular response to \ac{ER} stress, primarily induced by the accumulation of misfolded proteins within the \ac{ER}.\cite{lekstrom-himes1998biological,ghosh2012chop,oyadomari2004roles}
A key component of the \ac{UPR} is the \ac{CHOP}, which is upregulated in response to prolonged \ac{ER} stress and plays a critical role in the initiation of apoptosis.\cite{vij2008chop,ghosh2012chop,goodall2010endoplasmic}
This pathway becomes particularly relevant in pathological conditions such as cancer, where \ac{ER} stress and \ac{UPR} dysregulation are often observed.\cite{flaherty2014discovery,fawcett2006introduction,oyadomari2004roles}

In this study, the authors performed a systematic \ac{HTS} of a diverse chemical library to identify molecules that specifically induce the \ac{CHOP} pathway. 
Multiple screens were performed and curated, and also published on PubChem.\cite{flaherty2014discovery,kim2023pubchem}
This screening led to the discovery of a class of sulfonamidebenzamide compounds that effectively activate the \ac{CHOP} pathway. 
Subsequent investigations, including structure--activity relationship studies, allowed these compounds to be optimised for improved potency and selectivity.\cite{flaherty2014discovery}

The work of \citeauthor{flaherty2014discovery} serves as an exemplary demonstration of the effectiveness of the \textsc{TwinBooster} pipeline in practical scenarios. 
This research is particularly valuable as it includes both primary and confirmatory \ac{CHOP} \acp{HTS}, and all associated data are publicly available. 
Crucially, this publication is not included in the ChEMBL database, ensuring its complete exclusion from the FS-Mol dataset and consequently from the training data used.\cite{mendez2019chembl}
This clear separation underlines the suitability of this case study to illustrate the zero-shot learning capabilities of the \textsc{TwinBooster} pipeline.

The molecular and textual information from the primary screen (\ac{AID} 2732) is extracted and then predictions are made by \textsc{TwinBooster}.
Zero-shot predictions aim to predict a single endpoint for an unseen task.

\begin{figure}[!thb]
     \centering
     \begin{subfigure}[b]{0.49\textwidth}
         \centering
         \includegraphics[width=\textwidth]{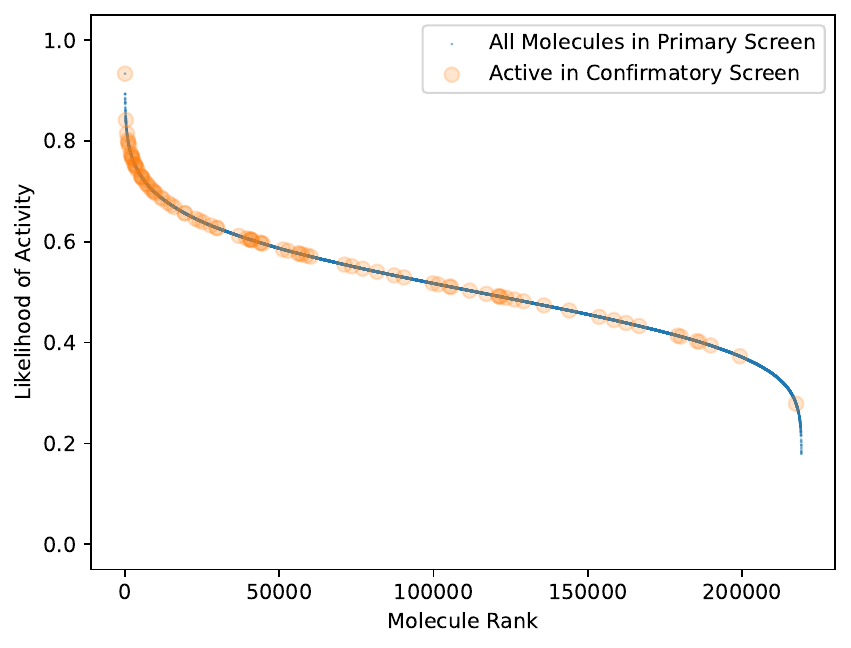}
         \caption{Prediction ranking}
         \label{fig:case_study_dist}
     \end{subfigure}
     \hfill
     \begin{subfigure}[b]{0.49\textwidth}
         \centering
         \includegraphics[width=\textwidth]{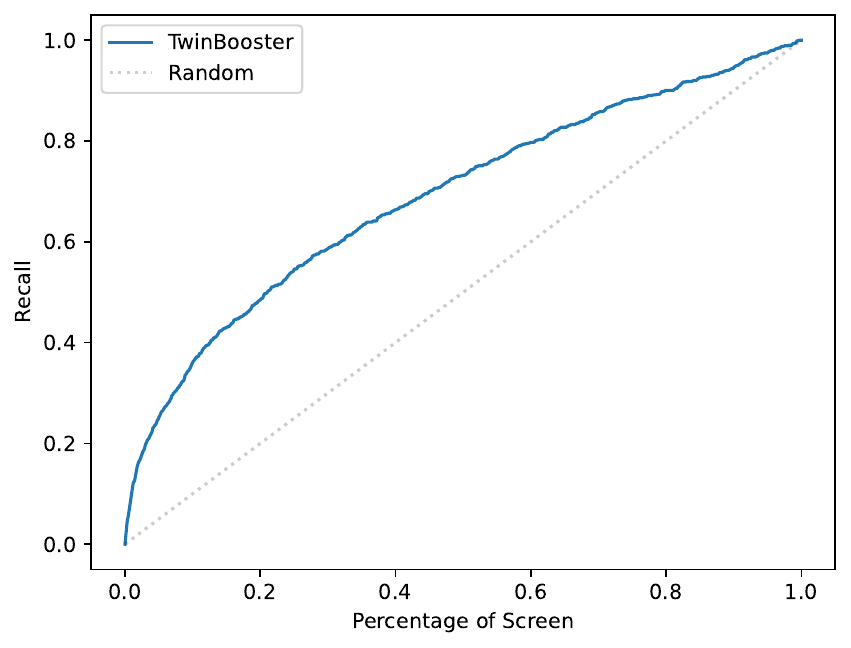}
         \caption{Recall}
         \label{fig:case_study_recall}
     \end{subfigure}
    \caption{Zero-shot predictions on the primary screen of the case study. \textbf{(a)} Ranked molecules based on zero-shot prediction: highlighting earlier discovery of confirmatory screening hits. \textbf{(b)} Recall curve: retrieved active compounds as a percentage of all active compounds based on the percentage of the primary screen provided.}
    \label{fig:case_study}
\end{figure}

\textsc{TwinBooster} correctly prioritises the majority of hits. 
It is highlighted in \cref{fig:case_study}, which shows its zero-shot predictions for approximately \num{220000} molecules for activities related to \ac{CHOP} and the \ac{UPR} pathway. 
The ability of the model to accurately classify these molecules, particularly in the upper likelihood range, is demonstrated in \cref{fig:case_study_dist}, where a notable enrichment in the discovery of confirmatory screening hits is observed among the actively classified molecules. 
Looking at the performance metrics of the primary screen in \cref{tab:cp_case_study}, the use of conformal predictions can relatively improve the \ac{DPR_AUC} of the model by up to \SI{72}{\percent}.
The proportion of confident predictions in this zero-shot case study bioassay is \SI{23}{\percent}.
Using this in the model during pre-screening can increase the overall hit rate and lead to a higher proportion of active molecules showing the desired biological effects.

\begin{table}[!htb]
    \centering
    \caption{Performance metrics of \textsc{TwinBooster} on the primary screen with or without conformal prediction. The confidence level is set to $\epsilon = 0.80$.}
    \label{tab:cp_case_study}
    \sisetup{
        table-alignment-mode = none,
        table-number-alignment = center,
        table-auto-round
    }
    \begin{tabular}{rSS}
    \toprule
    \multirow{2}{*}{\textbf{\textsc{TwinBooster}}} & \multicolumn{2}{c}{\textbf{Conformal Prediction}} \\
    & \xmark & \cmark \\
    \midrule
    \acs{ROC_AUC} (\si{\percent}) & 58.82 & 62.02 \\
    \acs{PR_AUC} (\si{\percent}) & 7.10 & 11.64 \\
    \acs{DPR_AUC} (\si{\percent}) & 3.34 & 5.73 \\
    \bottomrule
    \end{tabular}
\end{table}

Furthermore, results demonstrate the decent efficiency of the model: using only a \SI{20}{\percent} subset of the screening data, it is possible to accurately identify \SI{49}{\percent} of all active compounds (refer to \cref{fig:case_study_recall}). 
This efficiency is further highlighted when the provided data is increased to \SI{50}{\percent}, at which point approximately \SI{73}{\percent} of active molecules are correctly identified. 
This finding highlights the potential of the model to streamline the screening process by requiring significantly less molecules to achieve meaningful results.
Recall is a crucial metric in this context as it measures the ability of the model to correctly identify all relevant instances (in this case active molecules). 

\begin{figure}[!thb]
    \centering
    \includegraphics[width=.5\textwidth]{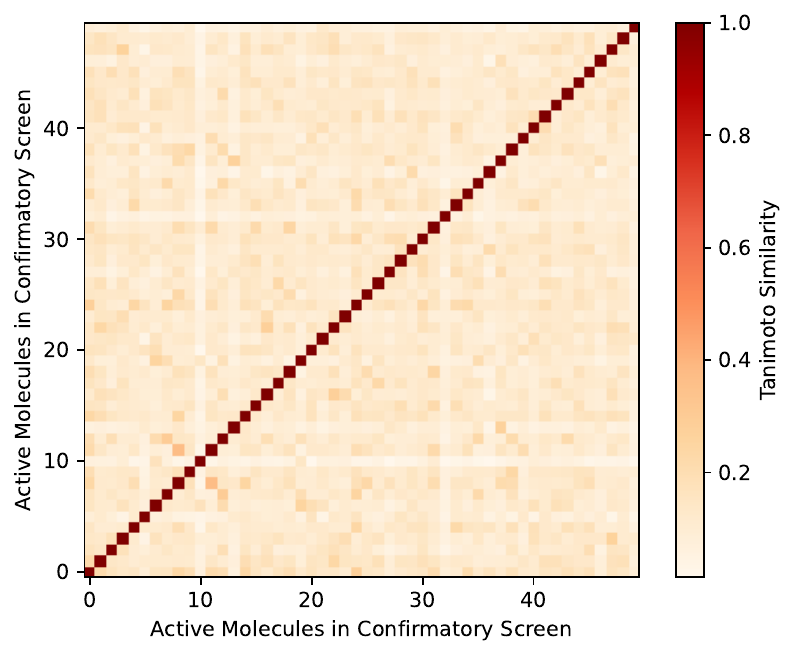}
    \caption{Tanimoto similarity of active compounds from the confirmatory screen. Shown is a random selection of 50 compounds, to highlight the structural similarities and differences.}
    \label{fig:scaffold_diversity}
\end{figure}

Furthermore, this study includes a systematic evaluation of whether the observed results can be attributed to the structural similarity of the compounds. 
For this purpose, the Tanimoto similarity is calculated as shown in \cref{fig:scaffold_diversity}.
This analysis reveals a significantly diverse structural distribution among the compounds, suggesting that similarities are not driving the observed performance. 
In \cref{fig:murcko_scaffold_diversity} the number of unique Murcko scaffolds from confirmatory screen relative to percentage from primary screen is shown.
\textsc{TwinBooster} is capable to reach an area under the curve of \SI{60.77}{\percent} compared to \SI{50}{\percent} in case of a random selection, i.e. with a \SI{25}{\percent} subset of the screening data it is possible to capture \SI{>39}{\percent} of unique Murcko scaffolds.\cite{landrum2020rdkit}
The ability of \textsc{TwinBooster} to discriminate and identify a wide range of potentially active compounds highlights its utility in streamlining the drug discovery process, paving the way for more targeted and expedient identification of promising compounds.

\section{Conclusion and Outlook}
In this study, we have taken a step towards improving the capabilities of drug discovery and development. 
By integrating a fine-tuned \ac{LLM} with the Barlow Twins architecture, and further employing \acp{GBM} for training and prediction, our zero-shot \textsc{TwinBooster} framework represents a novel approach to molecular property prediction, particularly in scenarios where data is scarce.\cite{zbontar2021barlow}

The effectiveness of \textsc{TwinBooster} in zero-shot learning tasks, as evidenced by its performance on the FS-Mol benchmark as well as in a \ac{HTS} case study, suggests that this methodology could be an important tool in the early stages of drug discovery.
It is able to outperform the best performing few-shot baseline at 16 support molecules provided by FS-Mol.\cite{stanley2021fsmol}

In addition, \textsc{TwinBooster}'s prediction model can help improving efficiency and economics of drug discovery. 
By diminishing the number of molecules that require empirical screening, it accelerates the research timeline and cuts down both time and associated costs, thereby enhancing the overall efficiency and cost-effectiveness of drug development. 

However, it is important to recognise the complexity of predicting very different assays in comparison to the training data in a zero-shot scenario.
While the results are promising, they represent one step in an ongoing journey of scientific exploration and innovation.

\section*{Data and Code Availability}

The system used for computational work has an AMD Ryzen Threadripper PRO 5995WX \ac{CPU} with 64/128 cores/threads with \SI{1024}{\giga\byte} \ac{RAM}.
Additionally, the server is equipped with a NVIDIA RTX 4090 \ac{GPU} with \SI{24}{\giga\byte} V\ac{RAM}.

The fine-tuned DeBERTa V3 model on the PubChem corpus is available on HuggingFace \url{https://huggingface.co/mschuh/PubChemDeBERTa}, as well as the augmented version \url{https://huggingface.co/mschuh/PubChemDeBERTa-augmented}.

As a Python package, it can be installed using \texttt{\$\:pip install twinbooster}.
The code is available on GitHub \url{https://github.com/maxischuh/TwinBooster}, where you can also find the model data.

\bibliography{ref.bib}

\printacronyms

\appendix

\section{Appendix -- Results}

\subsection{Zero-shot benchmark}

The zero-shot performances of 10 replicates are averaged and compared with \acs{PN} from FS-Mol.\cite{stanley2021fsmol}
Only means per task are used for the Wilcoxon test as only 5 replicates were performed on the \acs{PN} at 16 support molecules from FS-Mol.
The Wilcoxon test yields $p \simeq 0.0478$, for \acs{PR_AUC} and \acs{DPR_AUC}.\cite{virtanen2020scipy} 
Significance is indicated at $p < \alpha$, where $\alpha = 0.05$.
In addition, only the \num{122} intersecting tasks (tasks that are present in both FS-Mol and PubChem and therefore have an assay description) are evaluated.

The \acp{AID} are \numlist[group-digits = none]{521;689;881;883;899;1215;1394;1540;1708;1750;2161;2230;2364;2572;31668;48288;52163;218702;310904;404304;449749;456868;463120;482894;485349;485367;488785;488789;488835;488921;493182;493248;504729;507074;507077;588344;588345;588811;602234;602235;602374;602386;652135;720021;720033;720044;720046;720076;720081;720107;720113;720115;720127;720130;720132;720134;720136;720137;720146;720157;720162;720163;720175;720180;720185;720189;720191;720200;720202;720207;720233;720237;720246;720248;720251;720261;720262;720267;720276;720278;720281;720285;720289;720290;720295;720298;720310;720312;720319;720323;720329;720331;720339;720354;720357;720359;720361;720370;720373;720384;720392;720395;720421;720422;720427;720439;720442;720445;720446;720450;720453;720463;720473;720478;720481;720482;1053173;1207589;1207591;1207592;1438147;1501337}.

Comparing zero-shot performances of 10 replicates with or without conformal prediction on FS-Mol yields $p \simeq 0.0020$, for \acs{ROC_AUC}, \acs{PR_AUC} and \acs{DPR_AUC} on a Wilcoxon test.\cite{virtanen2020scipy} 
Significance is indicated at $p < \alpha$, where $\alpha = 0.05$.

\subsection{Ablation study}

The \ac{LSA} model\cite{deerwester1990indexing} is pre-trained on the PubChem corpus. 
This process is carried out in a manner similar to the approach used in \citeauthor{seidl2023enhancing}, using the Python packages \texttt{TfidfVectorizer} and \texttt{TruncatedSVD}.\cite{seidl2023enhancing,scikit-learn}
The resulting text embeddings for each bioassay have a dimensionality of 355.

In parallel, the hyperparameter optimisation for the Gradient Boosting Machine (GBM) is performed in a similar way as described in \cref{sec:gbm_hp}.

The $p$-values of the statistical tests using the Wilcoxon test with Bonferroni correction are provided in \cref{tab:p_val_ablations}.\cite{virtanen2020scipy,bonferroni1936teoria}

\begin{table}[!htb]
    \centering
    \caption{All $p$-value results are tested pairwise using the Wilcoxon test with Bonferroni correction of different ablation experiments on FS-Mol.\cite{virtanen2020scipy,bonferroni1936teoria} 10 replicates each are performed. Significance is indicated at $p < \alpha$, where $\alpha = \frac{0.05}{3}$. If $p \geq \alpha$, values are in italics.}
    \label{tab:p_val_ablations}
    \begin{adjustbox}{max width=\textwidth}
    \begin{tabular}{rccc}
    \toprule
     & \textbf{\textsc{TwinBooster}} & \textbf{\textsc{TwinBooster}} & \textbf{\text{ECFP + PubChemDeBERTa}} \\[-0.6ex]
    $p$-values & {\footnotesize \textbf{vs.}} & {\footnotesize \textbf{vs.}} & {\footnotesize \textbf{vs.}} \\[-0.4ex]
     & \textbf{\text{ECFP + PubChemDeBERTa}} & \textbf{\text{ECFP + LSA}} & \textbf{\text{ECFP + LSA}} \\
    \midrule
    \acs{ROC_AUC} & \itshape 0.1055 & 0.0020 & 0.0020 \\
    \acs{PR_AUC} & 0.0059 & 0.0020 & 0.0020 \\
    \acs{DPR_AUC} & 0.0059 & 0.0020 & 0.0020 \\
    \bottomrule
    \end{tabular}
    \end{adjustbox}
\end{table}

\subsection{Case study}

\begin{figure}[!ht]
    \centering
    \includegraphics[width=.5\textwidth]{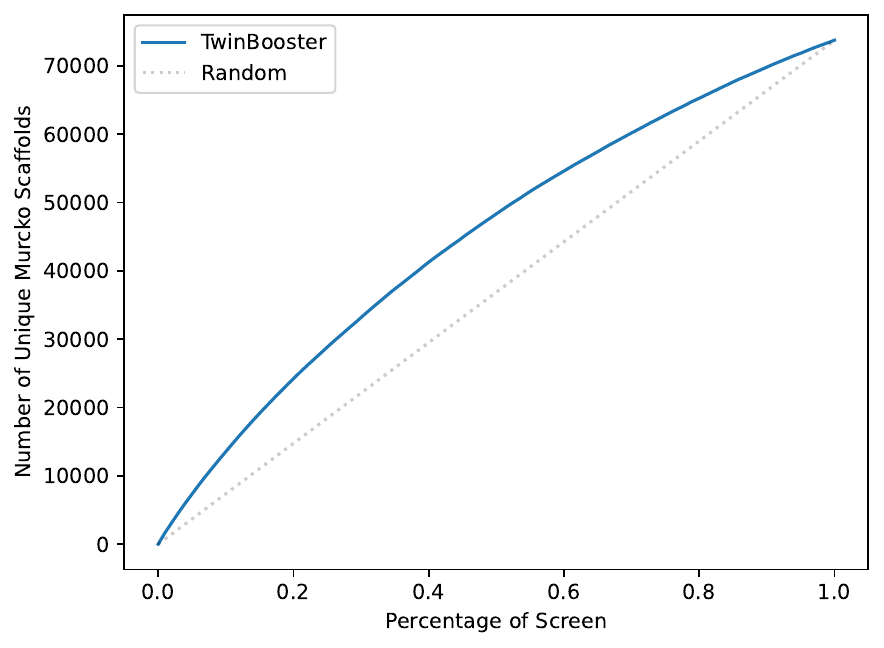}
    \caption{Number of detected unique Murcko scaffolds in relation to the percentage of scaffolds from the primary screen.}
    \label{fig:murcko_scaffold_diversity}
\end{figure}

\Cref{fig:murcko_scaffold_diversity} shows an enrichment of the number of unique Murcko scaffolds identified by \textsc{TwinBooster} and their proportional representation from the primary screen. 
This indicates that \textsc{TwinBooster} does not prioritise certain scaffolds, but enriches the scaffold diversity compared to random scaffold selection.
Highlighting the importance of scaffold variability in early drug development, this increase in diversity is critical to the identification of potential leads.

\paragraph{Primary screen} This represents title, description and protocol used for the zero-shot prediction of the primary screen (\ac{AID} 2732\footnote{\url{https://pubchem.ncbi.nlm.nih.gov/bioassay/2732}}), based of the PubChem entry:\cite{kim2023pubchem}

\blockquote{HTS for small molecule inhibitors of CHOP to regulate the unfolded protein response to ER stress.
Many genetic and environmental diseases result from defective protein folding within the secretory pathway so that aberrantly folded proteins are recognized by the cellular surveillance system and retained within the endoplasmic reticulum (ER). Under conditions of malfolded protein accumulation, the cell activates the Unfolded Protein Response (UPR) to clear the malfolded proteins, and if unsuccessful, initiates a cell death response. Preliminary studies have shown that CHOP is a crucial factor in the apoptotic arm of the UPR; XBP1 activates genes encoding ER protein chaperones and thereby mediates the adaptive UPR response to increase clearance of malfolded proteins. Inhibition of CHOP is hypothesized to enhance survival by preventing UPR programmed cell death. There are currently no known small molecule CHOP inhibitors either for laboratory or clinical use.
To identify small molecule inhibitors of the UPR pathway, mediated by CHOP, a cell-based luciferase reporter assay using stably transfected CHO-K1 cells with luciferase driven by the CHOP promoter has been developed. The assay have been optimized and validated in 384-well format and used to screen for inhibitors of tunicamycin-induced CHOP in HTS. These identified compounds will have potential therapeutic application to diverse disease states ranging from diabetes, Alzheimer's disease, and Parkinson's disease, to hemophilia, lysosomal storage diseases, and alpha-1 antitrypsin deficiency.
Reagents:
1. Cell line: CHO-CHOP cells with a luciferase reporter driven by the CHOP promoter (provided by assay PI)
2. Cell growth media (Ham's F12 + Glutamax, 10\% FBS, 1X non-essential amino acids, and penicillin:streptomycin) (Invitrogen)
3. Tunicamycin (Calbiochem)
4. SteadyGlo reagent (Promega)
Protocol:
1. 40 uL of medium containing CHO-CHOP cells (3000-4000) were dispensed to 384 well white opaque plates (Corning \#3570) using a Multidrop combi (Thermo-Fisher Scientific). Plates were then incubated for 24 hrs at 37 degrees C, 5\% CO2.
2. 0.5 uL of library compounds (1 mM in DMSO) was added to wells using Sciclone (Caliper LifeSciences). The final concentration of compound is 10 uM.
3. 10 uL of fresh medium containing tunicamycin (Tm) (2.0 ug/ml, final concentration,) was then added and the plates were incubated for 15-18 hrs.
4. Medium was aspirated with an Elx405 plate washer (BioTek), leaving 10 uL of medium in the well. 10 uL of Steady-Glo was added to each well using a multildrop combi.
5. Luminescence signal was measured on an Envision Multilable plate reader (PerkinElmer).}

\end{document}